\documentclass[aps,pra,reprint,superscriptaddress,floatfix]{revtex4-2}

\usepackage{placeins}
\usepackage{graphicx}
\usepackage{amsmath}
\usepackage{braket}
\usepackage{xcolor}
\usepackage[normalem]{ulem}
\usepackage{siunitx}
\usepackage{multirow}
\usepackage{dcolumn}
\usepackage{notes2bib}
\bibnotesetup{
note-name = ,
use-sort-key = false
}
\newcommand{\expect}[1]{\langle#1\rangle}
\newcommand{\Tr}{\mathrm{Tr}}

\usepackage{hyperref}
\hypersetup{
    colorlinks=true,
    citecolor=teal,
    linkcolor=teal,
    filecolor=blue,      
    urlcolor=teal,
}
\usepackage[nameinlink,capitalise]{cleveref}

\usepackage{orcidlink}

\begin{document}

\title{Structural complexity of an SU(3) Fermi Hubbard model}

\author{Jiani Fu\,\orcidlink{0009-0008-7547-3220}}
\affiliation{Department of Physics, Harvey Mudd College, 301 Platt Blvd, Claremont, California 91711, USA}
\author{Zewen Zhang\,\orcidlink{0000-0003-2258-613X}}
\affiliation{Shanghai Qi Zhi Institute, Shanghai 200232, PR China}
\author{Eduardo Ibarra-Garc\'ia-Padilla\,\orcidlink{0000-0001-9165-0444}}
\email[]{eibarragp@g.hmc.edu}
\affiliation{Department of Physics, Harvey Mudd College, 301 Platt Blvd, Claremont, California 91711, USA}
\date{\today}

\begin{abstract}
Two-dimensional quantum gas microscopy provides an unparalleled tool to study quantum many-body systems using ultracold atoms. For the SU(2) Fermi Hubbard model (FHM), access to spin-resolved projective measurements has been vital for quantifying correlation functions and mapping out the phase diagram. Recent progress in quantum gas microscopy for experiments with ultracold alkaline-earth atoms, which are well described by the SU(N) FHM and are predicted to host exotic ground-state phases, calls for the development of theory-free numerical techniques to extract physical information from their projective measurements. To that end, we evaluate the multiscale structural complexity of snapshots of an SU(3) FHM in the square lattice at $1/3$-filling. We employ mean-field theory to generate spin-resolved density distributions and compute their structural complexity using rectangular coarse-graining windows. We demonstrate that these complexities are linked to relevant physical observables such as the entanglement entropy, and are extremely sensitive for locating phase boundaries. The results presented here validate the structural complexity as an efficient and reliable tool for analyzing the outputs of SU(N) quantum gas microscopes, offering a theory-free property, immediately accessible to experiments.
\end{abstract}

\maketitle

\section{Introduction}

Ultracold atoms in optical lattices (OLs) as quantum simulators, are a valuable tool to study strongly correlated quantum many-body systems~\cite{Altman2021,Topology_review}. These analog quantum simulators provide a powerful platform to investigate unresolved emergent phenomena by leveraging precise control over the the strength and range of interatomic interactions, the lattice geometry, and the filling fraction~\cite{lewenstein2012ultracold,choi2023quantum,Bloch2012,Gross2017,Schafer2020}. Notably, one of the major accomplishments of ultracold atoms in OLs is the precise engineering and exploration of the SU(2) Fermi-Hubbard model (FHM) in regimes that are difficult or inaccessible to simulate numerically~\cite{Bohrdt2021,Xu2025}.

In lower dimensions, quantum gas microscopy has revolutionized our ability to probe the FHM. In particular, spin-resolved projective measurements (or snapshots) have played a crucial role in quantifying non-local and long-range correlation functions~\cite{Gross_Bakr_review}. These findings have provided profound insights into fundamental problems in quantum magnetism~\cite{Parsons2016,Cheuk2016,mazurenko2017cold,Xu2022,Xu2025,Lebrat2024,Prichard2024,Mongkolkiattichai2023}, pseudogap physics~\cite{Brown2020,Chalopin2026,Kendrick2025}, and transport properties in strongly correlated systems~\cite{Brown2018,Nichols2019,GuardadoSanchez2020}. 

Recently, quantum gas microscopy has been extended to experiments with ultracold alkaline-earth atoms (AEAs) in OLs~\cite{Boub2024,Gas2026}. These experiments exploit the intrinsic SU(N) nuclear spin symmetry of $^{173}$Yb and $^{87}$Sr to engineer the SU(N) FHM with tunable $N\leq10$~\cite{Gorshkov2010,Cazalilla2014,Boub2024,Gas2026,Hofrichter2016,Taie2012,Ozawa2018,Taie2022,Tusi2022,Pasqualetti2024,He_2019,Takahashi_review,EIGP_SC_2025}. Investigation of SU(N) FHMs has gained interest because they are predicted to host rich phase diagrams~\cite{Honerkamp2004,Toth2010,Corboz2011,Nataf2014,Bauer2012,Assaad2005,Zewen2026,Feng2023,IbarraGP2023,Bird2025,Rajiv2022_2,Hingorani2022,Sotnikov2014,Sotnikov2015,Sotnikov2020,Romen2020,Nie2017,Hafez2018,Hafez2019,Hafez2020,Goubeva2017,Gorelik2009,Titvinidze2011,Stepp2026,PerezRomero2021,Yamamoto2020} and are relevant for describing multiorbital materials, transition metal-oxides~\cite{Tokura2000,Dagotto2001,Li1998}, orbitally selective Mott transitions~\cite{delRe2018,Medici2005,Medici2014,Zavatti2025,Dasgupta2025,Fujii2026}, and robust itinerant ferromagnetism~\cite{Katsura2013,Bobrow2018,Rajiv2022}.

While the rapid developments in quantum gas microscopy for AEAs in OLs are bringing us closer to understanding the physics of the SU(N) FHM, two challenges remain. First, most of the phase diagrams of these models remain unexplored because conventional numerical techniques, such as exact diagonalization (ED), numerical linked cluster expansions (NLCE), density matrix renormalization group (DMRG), and quantum Monte Carlo (QMC)~\cite{DMRG_White,DMRG_review,DQMC_1,DQMC_2,NLCE_review}, suffer from either exponential scaling or the fermion \textit{sign problem}~\cite{IbarraGP2021,Loh1989,Troyer2005,Iglovikov2015,Pan_signproblem}. Second, the plethora of predicted ground states exhibit a complicated $N$ dependence. These exotic ground states, including chiral spin liquids and topological phases~\cite{Hermele2011,Nataf2016,Chen2016, Hermele2009}, display complex spatial patterns and emerging competing orders that are difficult to describe quantitatively. Such difficulty calls for the development of versatile numerical techniques to extract physical information from spin-resolved projective measurements without relying on a priori theoretical assumption.

One powerful route to identify phase boundaries and extract physical information from projective measurements in an unbiased way, is by harnessing machine learning and other data-driven techniques~\cite{Kelvin2017,Bohrdt2019,Khatami2020,Striegel2023,Johnston2022}. Consequently, in this work we leverage the unbiased measure named the multiscale structural complexity~\cite{Bagrov2020}. This measure quantifies pattern dissimilarity across different scales~\cite{Wolpert2007}, and uses concepts inspired by the renormalization-group flow to aggregate information about different scale correlations present in the system. To date, the structural complexity measure has been successfully used to analyze snapshots of one- and two-dimensional quantum and classical systems~\cite{Sotnikov2022,Mazurenko2023,IbarraSC2024}, as well as systems out of equilibrium~\cite{Maletskii2024}. Notably, this metric successfully identified off-diagonal bond-density-wave phase transitions in the half-filled extended FHM while relying solely on diagonal density snapshots~\cite{Xiao2024}. Furthermore, its behavior strongly correlates with entanglement entropy~\cite{Sotnikov2022,Sotnikov2024,Xiao2024}, highlighting its ability to capture non-local quantum correlations.

\begin{figure}[tbp!]
    \centering
    \includegraphics[width=\linewidth]{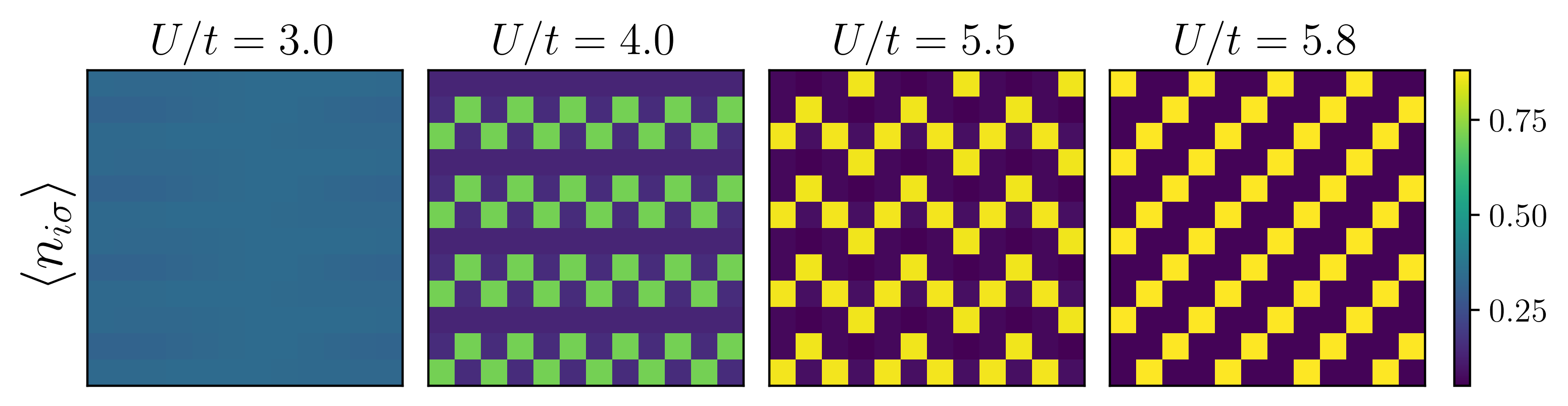}
    \caption{Examples of mean-field ground state density profiles $\expect{n_{i\sigma}}$ at $1/3$-filling for selected values of $U/t$. Results for the other two spin flavors (not shown) are related by translations. At $U/t=3.0$, $\expect{n_{i\sigma}}$ is spatially uniform and consistent with a metallic phase. As the interaction strength increases, the tooth phase is observed ($U/t=4.0$), then the zig-zag phase ($U/t=5.5$), and finally the stripe phase ($U/t=5.8$).}
    \label{fig:number_density}
\end{figure}

In this work, we employ the multiscale structural complexity to study the SU(3) FHM in the square lattice at $1/3$-filling. This model is predicted to display a quantum critical point signaling the metal-to-insulator transition for $U_c>0$, and a series of antiferromagnetic (AFM) phases at $U>U_c$~\cite{Feng2023,IbarraGP2023,Bird2025,Zewen2026}. In the present study, we generate spin-resolved density distributions using mean-field theory (MFT) and compute their complexity measure using rectangular coarse-graining windows. We demonstrate that this metric strongly correlates to the entanglement entropy and accurately locates the phase boundaries of the mean-field (MF) diagram of the SU(3) FHM. We benchmark our findings against those from Ref.~\cite{Zewen2026}, which observed a metallic phase for $U/t<3.50$, and three ordered AFM phases (see Fig.~\ref{fig:number_density}): the \textit{tooth} [$U/t\in(3.50 ,4.75)$], the \textit{zig-zag} [$U/t\in(4.75,5.65)$], and the \textit{stripe} [$U/t > 5.65$], with ordering wavevectors $\mathbf{q} = (2\pi/3,\pi),(2\pi/3,\pi/2)$, and $(2\pi/3,2\pi/3)$, respectively.

The remainder of this paper is organized as follows. In Sec.~\ref{sec::Model_Methods} we present the model we study, the observables we measure, and the methods used. In Sec.~\ref{sec::Results} we present our main findings. In Sec.~\ref{sec::Conc} we present our conclusions.

\section{Model and methods}\label{sec::Model_Methods} 

\subsection{SU(3) Fermi-Hubbard model}
We study the two-dimensional SU(3) FHM, described by the Hamiltonian
\begin{equation}\label{eq:Hubbard_N1}
\hat{H} = -t \sum_{\langle i,j \rangle, \sigma} \left( \hat{c}_{i \sigma}^\dagger \hat{c}_{j \sigma}^{\phantom{\dagger}} 
+ \mathrm{h.c.} \right) + \frac{U}{2} \sum_{i,\sigma \neq \tau} \hat{n}_{i \sigma} \hat{n}_{i \tau}  - \sum_{i,\sigma} \mu_\sigma \hat{n}_{i \sigma},
\end{equation} 
where $\hat{c}_{i \sigma}^\dagger$ ($\hat{c}_{i \sigma}^{\phantom{\dagger}} $) is the creation (annihilation) operator for a fermion with spin flavor $\sigma$ on site $i$, $\hat{n}_{i \sigma} = \hat{c}_{i \sigma}^\dagger \hat{c}_{i \sigma}^{\phantom{\dagger}}$ is the number operator for spin $\sigma$ on site $i$, $\expect{i,j}$ denotes sum over nearest neighbors, $t$ is the nearest-neighbor hopping amplitude, $U$ is the interaction strength, and $\mu_\sigma$ is the chemical potential that controls the fermion density of spin $\sigma$ in the grand canonical ensemble. We consider the repulsive case $U>0$ at $1/3$-filling $(\expect{n_\sigma}=1/3)$, and set the energy scale to be $t=1$. We consider 2D rectangular lattices with $N = L_x \times L_y$ sites and periodic boundary conditions. The linear dimensions are set to $L_x=24$ and $L_y=12$, unless otherwise specified.

\subsection{Mean-field theory}
To get the ground states of Hamiltonian~\eqref{eq:Hubbard_N1}, the Hartree approximation is used~\cite{Scholle2023}. This approximation expands the number operator around its mean-field value, $n_{i\sigma}=\braket{n_{i\sigma}} + \delta n_{i\sigma}$. Truncating fluctuations at the first order gives
\begin{align}\label{eq::Hartree_H}
    H = &
    -t\sum_{\langle i,j\rangle,\sigma} \hat{c}^\dagger_{i,\sigma}\hat{c}_{j\sigma} +\frac{U}{2}\sum_{i,\sigma\neq\tau}
    (\hat{n}_{i\sigma} \braket{ n_{i\tau}}+\braket{n_{i\sigma}} \hat{n}_{i\tau}) \nonumber \\
    &-\frac{U}{2}\sum_{i,\sigma\neq\tau}\braket{n_{i\sigma}} \braket{ n_{i\tau}}
    - \sum_{i,\sigma}  \mu_\sigma \hat{n}_{i \sigma}.
\end{align}

The approximation gives an effective non-interacting model in the presence of an external field. This field, together with the chemical potential, is solved iteratively to converge to a density distribution $\braket{n_{i\sigma}}$. The self-consistent solution with lowest energy is taken as the mean-field ground state. We consider the convergence condition of $\sum_{i,\sigma} \lvert \expect{n_{i\sigma}}_{\text{iter}=p} - \expect{n_{i\sigma}}_{\text{iter}=p+5} \rvert < \epsilon$ at the $p^\text{th}$ iteration~\cite{Zewen2026}, and set $\epsilon = 1\times 10^{-7}$.

The resulting Hamiltonian decouples the fermion operators of the different spin flavors, each of which is quadratic in the fermion operators with eigenvalues $\{\epsilon_{k \sigma}\}$ and eigenfunctions $\phi_{ik}^{(\sigma)}$. The transformation ${c_{i\sigma} = \sum_k\phi_{ik}^{(\sigma)}c_{k\sigma}}$ diagonalizes eq.~\eqref{eq::Hartree_H}. The ground state is given by $\ket{\Psi_0} = \prod_{\sigma,k^*} c_{k\sigma}^\dagger \ket{0}$, with energy $E_{0} = \sum_\sigma \sum_{k^*} \epsilon_{k\sigma}$, where $k^*$ 
restricts the product to states satisfying $\epsilon_{k\sigma} \leq \mu_\sigma$, where the chemical potential $\mu_\sigma$ determines the single-particle filling for spin flavor $\sigma$.

\subsection{Entanglement entropy}

The bipartite entanglement entropy, a measure of the degree of entanglement between two subsystems, is a central metric for distinguishing quantum phases of matter and characterizing their phase transitions~\cite{Vidal2003,Calabrese_2004,Calabrese_review,Laflorencie_review,Emidio2024,Teemu2022,Osterloh2002}. Therefore, we compute the entanglement entropy
\begin{equation}
    \mathcal{S}_E (\ell) = -\Tr\left[\rho_\mathcal{A} \ln \rho_\mathcal{A} \right],
\end{equation}
where $\mathcal{A}$ is one of the sub-partitions of size $\ell$, and $\rho_\mathcal{A}$ is its reduced density matrix.

Because the Hamiltonian in eq.~\eqref{eq::Hartree_H} is quadratic in the fermion operators, we can calculate $\mathcal{S}_E (\ell)$ via the correlation matrix method~\cite{Chung2001,Peschel_2003,Peschel2009,Wang2017}. In this method, the correlation matrix for spin $\sigma$ is given by
\begin{equation}
    C_{ij}^{(\sigma)} = \expect{\Psi_0|c_{i\sigma}^\dagger c_{j\sigma}^{\phantom{\dagger}}|\Psi_0} = \sum_{k^*} \phi_{ik}^{*(\sigma)} \phi_{jk}^{(\sigma)},
 \end{equation}
where $i,j \in \mathcal{A}$, and the entanglement entropy is expressed as,
\begin{equation}
    S_E (\ell) = -\sum_{\sigma,m} \Big[ \lambda_m^{(\sigma)} \ln(\lambda_m^{(\sigma)}) + (1-\lambda_m^{(\sigma)}) \ln(1-\lambda_m^{(\sigma)}) \Big],
\end{equation}
where $\lambda_m^{(\sigma)}$ are the eigenvalues of the $C_{ij}^{(\sigma)}$ matrices.

\subsection{Structural complexity}

\begin{figure}[htbp!]
    \centering
    \includegraphics[width=\linewidth]{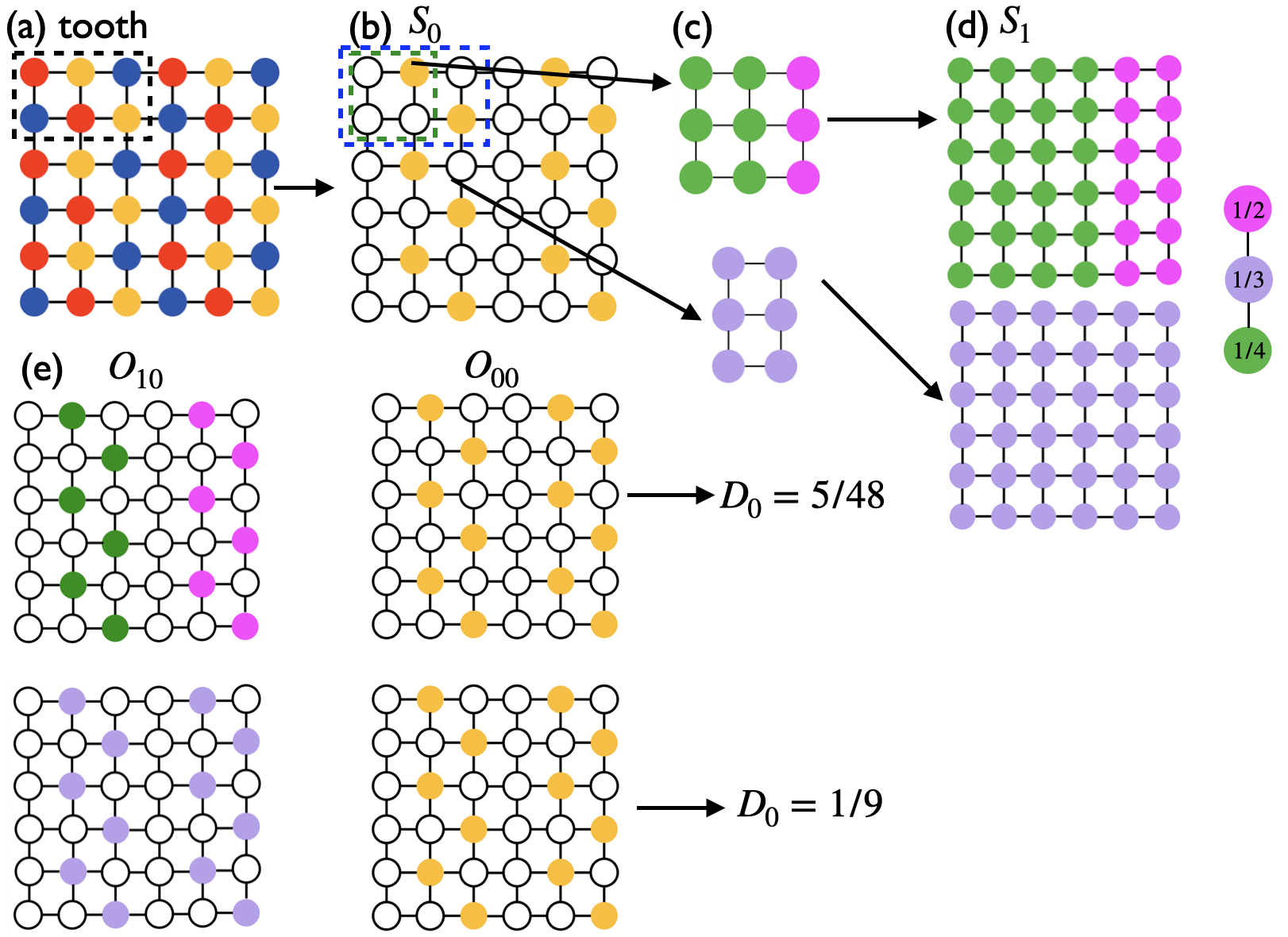}
    \caption{Structural complexity procedure. The classical spin arrangement of the tooth phase is depicted in (a), where the different colors denote the three possible spin flavors. The density distribution $S_0$ of one of the spin components (in this case, yellow), is coarse-grained using a $\Lambda_x \times \Lambda_y$ window in (b) [$2\times 2$ and $3\times 2$ windows are shown in green and blue dashed lines, respectively]. The coarse-grained images obtained using both windows are shown in (c). They are then resized to their original size in (d), denoted $S_1$. Overlaps $O_{i0}$ between images $S_i$ and $S_0$ are computed by performing an element to element product in (e). The first value of dissimilarity $D_0$ is obtained by averaging over sites of the quantity $D_0=|O_{10}-(O_{11}+O_{00})/2|$. This coarse-graining process is then repeated with $S_1$ to compute $D_1$, and so on.}
    \label{fig:example_cal}
\end{figure}

The multiscale structural complexity employs concepts inspired by the renormalization-group flow to quantify pattern dissimilarity across different length scales to extract a single number that characterizes an image~\cite{Bagrov2020}. This number is obtained via a series of coarse-graining steps to effectively aggregate information that appears at different scales in the system. Formally, the structural complexity $C_0$ is defined as:
\begin{align}
    C_0 &= \sum_{k=0}^{k_\mathrm{max} -1} D_k = \sum_{k=0}^{k_\mathrm{max}-1} \bigg\vert O_{k+1,k} - \frac{1}{2} \left( O_{k+1,k+1} + O_{k,k}\right) \bigg\vert,
    \label{eq:C0}
\end{align}
where $D_k$ is named the ``dissimilarity'' at coarse-graining step $k$, $k_\mathrm{max}$ is the total number of coarse-graining steps, and  
\begin{equation}
    O_{k,p} = \frac{1}{N} \sum_{i=1}^N s^{(k)}_i s^{(p)}_i 
\end{equation}
is the overlap function, where the image has $N$ pixels, and $s^{(k)}_i$ corresponds to the value of the pixel at site $i$ at coarse-graining step $k$. In this work we perform the coarse-graining procedure on rectangular windows of size $\Lambda_x\times\Lambda_y$. The choice of rectangular coarse-graining windows instead of the commonly used square ones allows us to extract further information from the same images, as we will evidence in the Results section. The coarse-graining procedure for $\Lambda_x\times\Lambda_y$ windows and the calculation of the overlaps is depicted in Fig.~\ref{fig:example_cal}.

In addition to $C_0$, we also define $C_1 = C_0 - D_0$, i.e. starting the sum in eq.~\eqref{eq:C0} at $k=1$ rather than $k=0$. As it has been shown previously~\cite{IbarraSC2024}, contributions to $C_0$ and $C_1$ fall off rather rapidly with $k$, so in practice only a limited number of terms need to be considered.

While the challenges of getting MFT to converge to a single, final set of self-consistent densities $\expect{n_{i\sigma}}$ are generally viewed as a limitation of the approach (linked to the appearance of many, closely spaced, ground states with possibly different types of order), in this work we turn that drawback into an advantage to quantify the structural complexity of the model. To do so, we first generate mean-field snapshots of density distributions $\expect{n_{i\sigma}}$ for fixed $U/t$ at the target filling for many different initial conditions (or random seeds). Then, we post-select the snapshots by only considering those solutions with an energy $E$ satisfying $E-E_\mathrm{min} \leq \Delta E_\mathrm{cut}$, where $E_\mathrm{min}$ is the overall minimum energy over the dataset at a given $U/t$. In addition to this energy cutoff, for $U/t \in [4.8,5.6]$, where the tooth and the zig-zag phases actively compete, i.e. their energies are closely spaced, we apply a more stringent filter, which we named \textit{phase filtering}. This process effectively rejects snapshots in the tooth phase, as these are at higher energy than those in the zig-zag phase (see Appendix~\ref{App::postselection} for further discussion on the post-selection procedure). We set $\Delta E_\mathrm{cut}/t= 4 \times 10^{-3}$, which ensures that the analyzed snapshots correspond to density distributions as close to the ground state as possible while having no effect on $C_0$ and eliminating spurious signals on $C_1$. We then apply point-group symmetries to increase the number of samples eightfold. 

The structural complexity is computed for one large image obtained by concatenating the post-selected images, since this concatenation (or tiling) procedure has proven to be crucial in capturing quantum fluctuations~\cite{Sotnikov2022,Xiao2024,IbarraSC2024}. In practice, we tile a region of space using at least 50 of them after post-selection. This allows us to perform the coarse-graining procedure up to a minimum of 6 times $(k_\mathrm{max} = 6)$ for all $U/t$ values for the $2\times2$ coarse-graining window, and up to 3 times ($k_\mathrm{max}=3$) for the other coarse-graining windows. We evaluate the structural complexity for all spin flavors, which by SU(3) symmetry should be identical at $1/3$-filling, and present their average and standard error of the mean (sem) in the Results section, unless otherwise specified.

\section{Results}\label{sec::Results}

\begin{figure}[tbp!]
    \centering
    \includegraphics[width=1.0\linewidth]{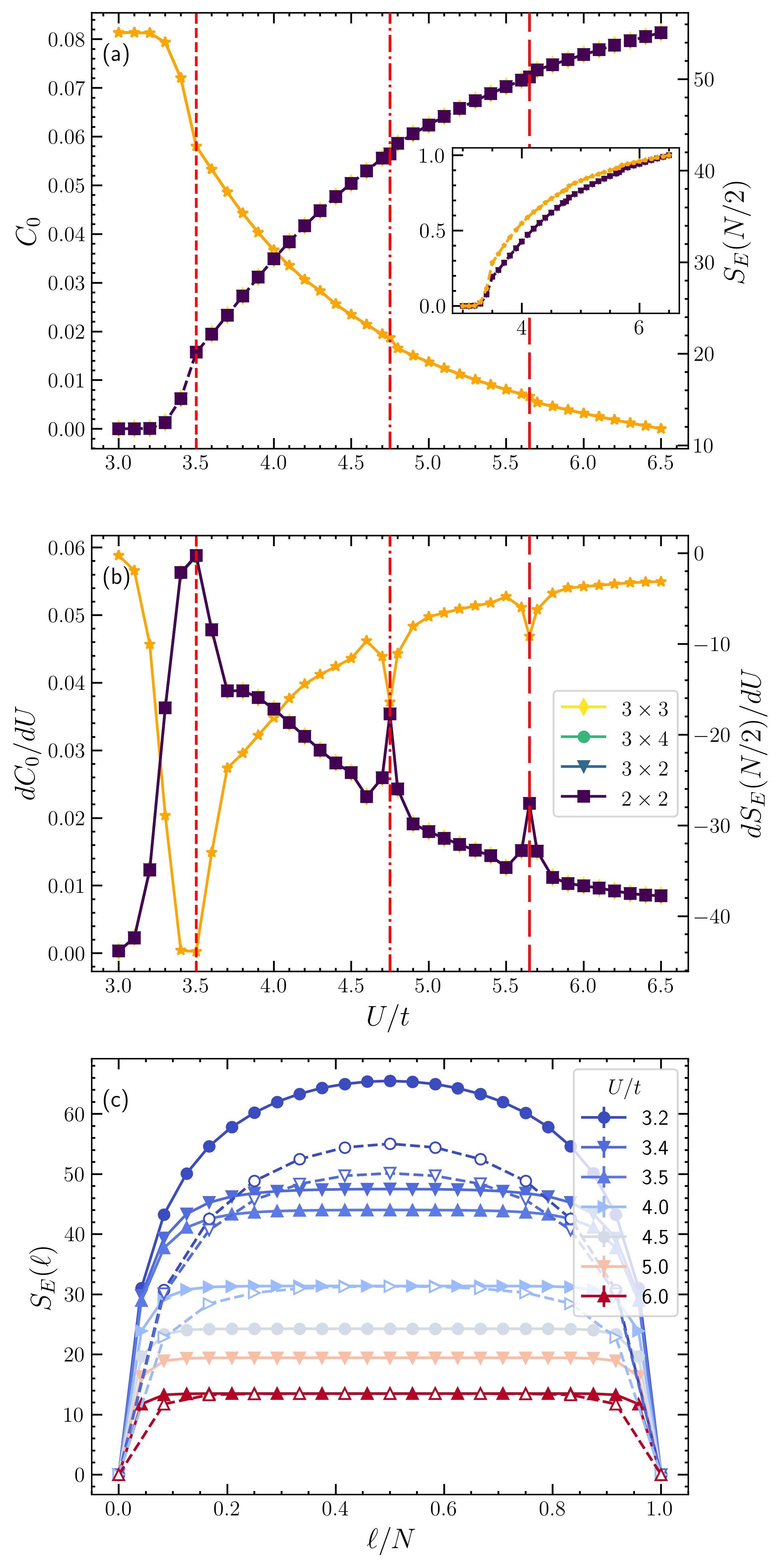}
    \caption{Results as functions of $U/t$ for different coarse-graining windows: (a) $C_0$ (left axis) and bipartite half-system entanglement entropy $S_E(N/2)$ (right axis, orange stars), and (b) $dC_0/dU$ (left axis) and $dS_E(N/2)/dU$ (right axis, orange stars). There is no visible difference between calculations using different windows (diamonds, circles, triangles, and squares cover each other up). Prior to taking the numerical derivatives, we smooth the observables using a moving average with a three-point window fitted with a local first-order polynomial (Savitzky-Golay filter). (c) Bipartite entanglement entropy $S_E(l)$ as a function of subsystem size $\ell$ for different values of $U/t$, computed for both $24\times12$ (open markers and dashed lines) and $24\times 24$ (solid markers) lattices. The inset in panel (a) corresponds to normalizing both quantities to unity and reflecting $S_E(N/2)$ at its midpoint. Red vertical lines in (a) and (b) correspond to the locations of the maxima in $dC_0/dU$.} 
    \label{fig:c0_c0diff_EE_4_10e4}
\end{figure}

In Fig.~\ref{fig:c0_c0diff_EE_4_10e4} we present the $U/t$ dependence of the structural complexity $C_0$ and the bipartite entanglement entropy $S_E(\ell)$. Figs.~\ref{fig:c0_c0diff_EE_4_10e4}~(a)-(b) evidence the strong correlation between structural complexity $C_0$ and the bipartite half-system entanglement entropy $S_E(N/2)$. Specifically, Fig.~\ref{fig:c0_c0diff_EE_4_10e4}(a) reveals an almost perfect proportional relationship between the two quantities. Normalizing both observables to unity and reflecting $S_E(N/2)$ about its midpoint produces a near-perfect collapse across $U/t$ [see the inset in Fig.~\ref{fig:c0_c0diff_EE_4_10e4}(a)]. Although $C_0$ and $S_E(N/2)$ do not perfectly line up for all values of $U/t$, their derivatives align extremely well [see Fig.~\ref{fig:c0_c0diff_EE_4_10e4}(b)], with $dC_0/dU$ capturing both the extrema and approximate slope of $dS_E(N/2)/dU$ across all values of the interaction strength. Moreover, Fig.~\ref{fig:c0_c0diff_EE_4_10e4}(b) demonstrates that the derivatives of $C_0$ and $S_E(N/2)$ accurately locate the phase boundaries of the model. Both derivatives exhibit extrema at $U/t=3.50, 4.75$ and $5.65$, which coincide with the phase boundaries predicted in Ref.~\cite{Zewen2026}. The latter are shown as vertical red lines. These extrema reflect the rapid reorganization of the ground-state density profiles as the system transitions through the distinct phases (see Fig.~\ref{fig:number_density}). 

The behavior of $C_0$ and $S_E(N/2)$ and their derivatives can be understood as follows. In the metallic phase, at $U/t <3.50$, where the density profile is spatially uniform, $S_E(N/2)$ is largest and $C_0$ is zero. As $U/t$ increases, the system approaches the metal-insulator transition at $U/t = 3.50$, with interactions opening a Mott gap, driving the formation of local moments and the appearance of the AFM tooth phase. This metal-insulator transition is further supported by the results in Fig.~\ref{fig:c0_c0diff_EE_4_10e4}(c), where the entanglement entropy is presented for subpartitions of the system in rows of length $L_x$ with $\ell$ sites. In the metallic phase the system is gapless and the entanglement entropy follows the Calabrese-Cardy formula~\cite{Calabrese_2004}
\begin{equation}
    S_E(\ell) = \frac{c}{3} \ln\left[\frac{L}{\pi a} \sin\left(\frac{\pi \ell}{L}\right)\right],
\end{equation}
where $L$ is the length of the system, $a$ is the lattice spacing, and $c$ is the central charge which is associated with the number of gapless excitation modes with a U(1) symmetry. As $U/t$ increases, the system enters a Mott insulating phase around $U/t=3.50$, opening a charge gap and thus $c\to 0$, leading to an almost independent dependence of $S_E$ on $\ell$. This explains why the metal-insulator transition [extrema at $U/t=3.50$ in Fig.~\ref{fig:c0_c0diff_EE_4_10e4}(b)] is accompanied by the sudden decrease in bipartite half-system entanglement entropy, and the rapid increase in the complexity.

For $U/t>3.50$, increasing $U/t$ favors stronger local moment formation (see Appendix~\ref{App:correlation}), leading to growth in $C_0$ alongside a reduction in $S_E(N/2)$. This inverse behavior persists across multiple regimes: first abrupt at the transition into the zig-zag phase [extrema at $U/t=4.75$ in Fig.~\ref{fig:c0_c0diff_EE_4_10e4}(b)], and then continuously beyond $U/t=4.75$ until reaching the boundary between the zig-zag and stripe phases [extrema at $U/t=5.65$ in Fig.~\ref{fig:c0_c0diff_EE_4_10e4}(b)].

Fig.~\ref{fig:c0_c0diff_EE_4_10e4}(c) also demonstrates that finite-size effects on $S_E(\ell)$ are negligible in the insulating phase, especially at $\ell = N/2$. While finite-size effects are larger in the metallic region, these do not modify the logarithmic shape observed in larger systems, and thus our conclusions remain.

\begin{figure}[tbp!]
    \centering
    \includegraphics[width=1.0\linewidth]{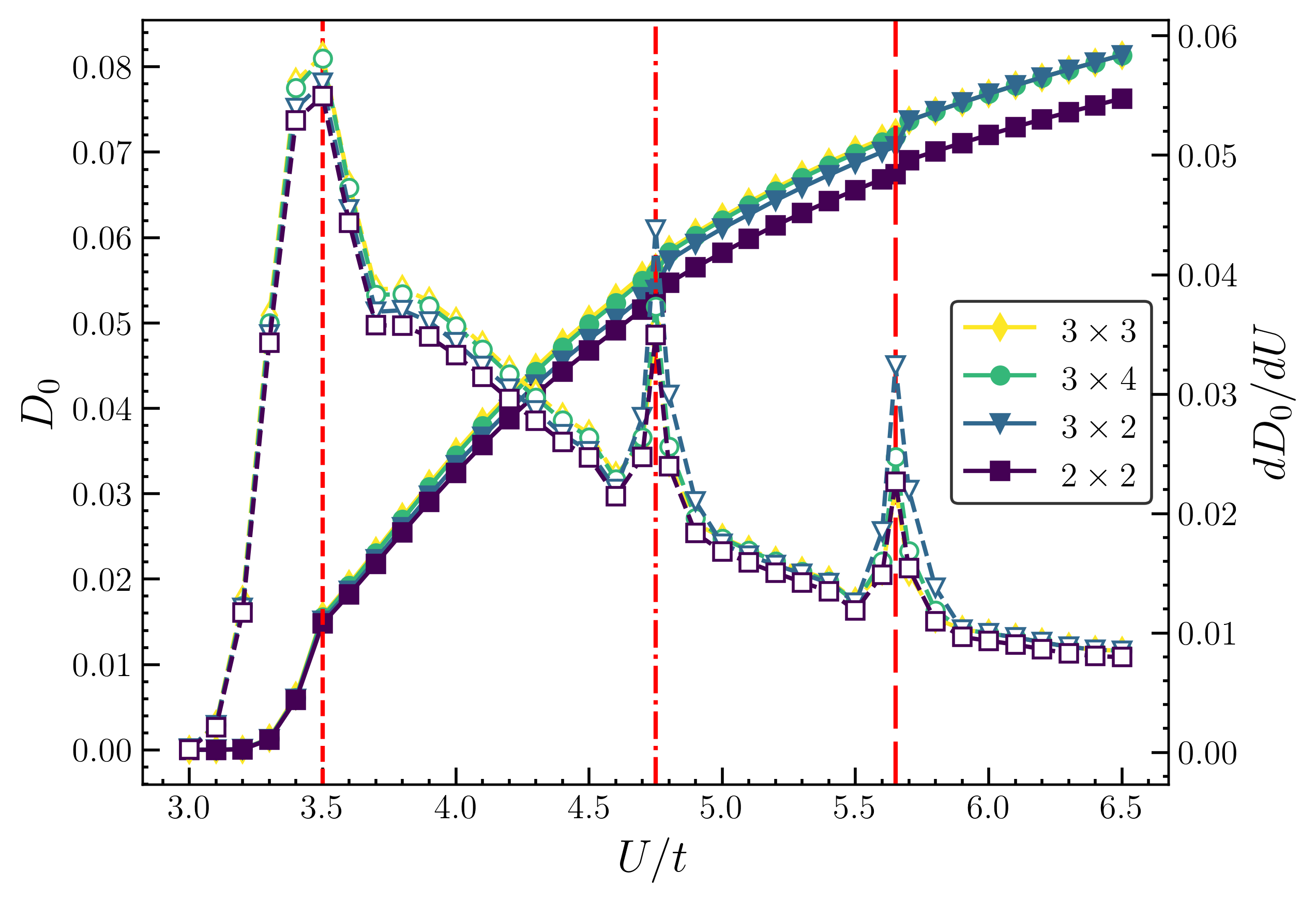}
    \caption{$D_0$ (left axis, solid markers) and $dD_0/dU$ (right axis, open markers) as functions of $U/t$ for different coarse-graining windows. We apply the same moving filter to the data prior to taking the numerical derivative. Red vertical lines are the same as those in Fig.~\ref{fig:c0_c0diff_EE_4_10e4}.}
    \label{fig:D0_plot}
\end{figure}

\begin{figure}[tbp!]
\centering
\includegraphics[width=0.95\linewidth]{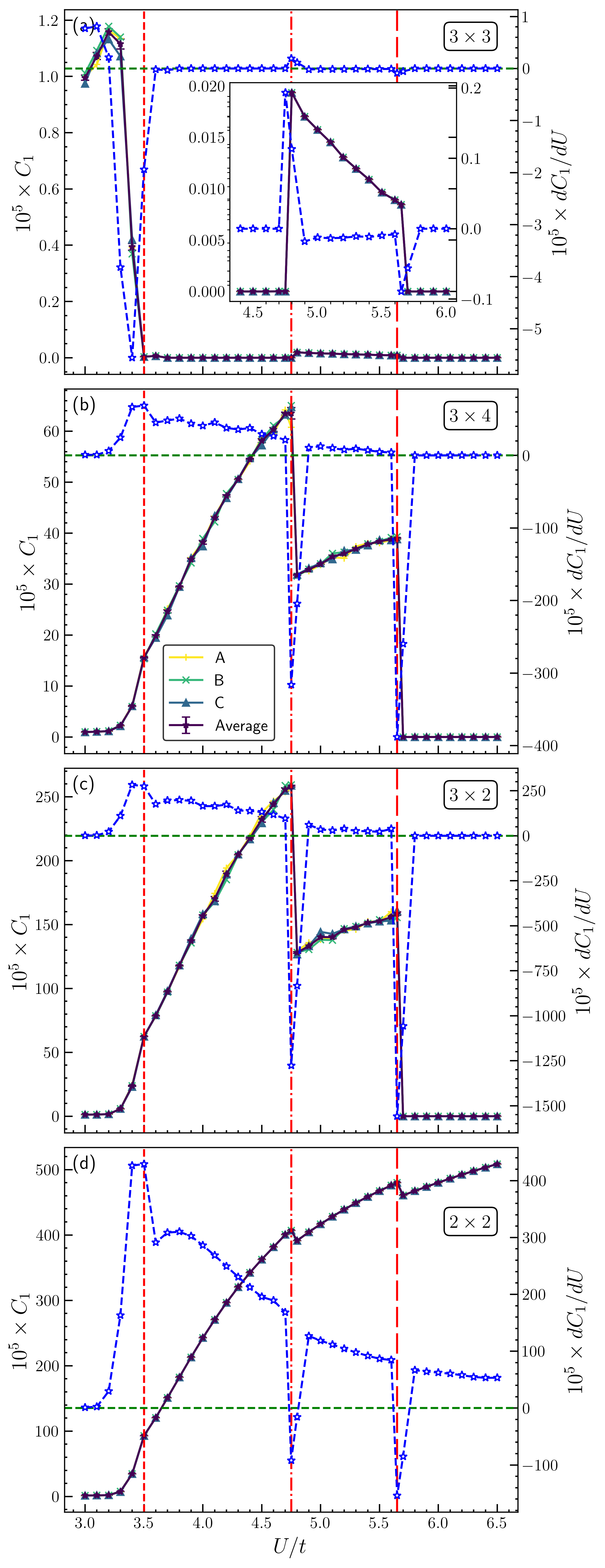}
 \caption{$C_1$ (left axis, solid markers) and $dC_1/dU$ (right axis, open blue markers) as functions of $U/t$ for different coarse-graining windows. We present $C_1$ for all spin flavors (A,B,C) as well as their average with error bars given by the sem. We present $dC_1/dU$ for the average only, to avoid clutter. Red vertical lines are the same as those in Fig.~\ref{fig:c0_c0diff_EE_4_10e4}.}
\label{fig:c1_12by24}
\end{figure}

Notably, the structural complexity $C_0$ is independent of the coarse-graining window used. This is shown in Figs.~\ref{fig:c0_c0diff_EE_4_10e4}(a)-(b), where we evaluate $C_0$ using $2\times2$, $3\times2$, $3\times4$, and $3\times3$ windows, and note that their markers cover each other up. Because of the insensitivity of $C_0$ to the coarse-graining window, it is instructive to analyze the behavior of its components: the dissimilarity after the first coarse-graining step $D_0$ (Fig.~\ref{fig:D0_plot}), and the complexity $C_1$, which captures the contributions from the remaining dissimilarities (Fig.~\ref{fig:c1_12by24}).

The behavior of $D_0$ closely follows the behavior of $C_0$ as a function of the interaction strength (see Fig.~\ref{fig:D0_plot}). The results show that the dominant contribution to $C_0$ arises from the first coarse-graining step, and that both derivatives show maxima at $U/t=3.50, 4.75$, and $5.65$, which mark the phase boundaries. However, the dependence of $D_0$ on the coarse-graining window reveals further information: In the metallic phase, $D_0$ has the same value for all windows, while in the AFM phases, $D_0$ is sensitive to the coarse-graining window. As $U/t$ increases past the metal-insulator transition, the $3\times3$ window gives the largest value of $D_0$, followed by the $3\times4$ and $3\times 2$ windows. This trend continues until the system enters the stripe phase at $U/t>5.65$, where those three windows give the same $D_0$. In contrast, the $2\times2$ window systematically yields a lower value than the rest of the windows for all $U/t>3.50$.

While $D_0$ is sensitive to local correlations within the initial window, $C_1$ aggregates the remaining contributions from higher dissimilarities, capturing the effects from correlations that are present at larger length scales. The dependence of $C_1$ on the coarse-graining windows is shown in Fig.~\ref{fig:c1_12by24}, along with its derivative as functions of $U/t$. First, we observe that in contrast to $D_0$, for $U/t>3.50$, the $2\times 2$ window yields the largest values of $C_1$, and the $3\times3$ window gives the smallest ones. Second, we discover that $C_1$ does not grow monotonically as $U/t$ increases, contrary to what is observed for $D_0$ and $C_0$. Third, we detect that, similar to $dC_0/dU$ and $dD_0/dU$, the extrema in $dC_1/dU$ identify the location of the phase boundaries. However, the sign of $dC_1/dU$ across the phase boundaries depends on the coarse-graining window used (see Table~\ref{table:C1_deriv}). 

To gain insight into the dependence of $D_0$ and $C_1$ on the coarse-graining windows and examine the role of quantum fluctuations on the complexity, it is instructive to analyze the classical spin configurations of the three AFM phases. In this analysis, we fill the lattice with the dominant spin flavor associated with classical configurations shown in Fig.~\ref{fig:number_density}. This is exemplified for the tooth phase in Fig.~\ref{fig:example_cal}(b) where for $S_0$ white circles take on the value of zero, and yellow circles the value of one. 

Under this approximation, it is easy to derive that for the three AFM phases $C_0=1/9$. Specifically, $D_0 = 1/9$ for the $3\times n$ ($n=2,3,4$) windows, and $D_0= 5/48$ for the $2\times2$ window (see Appendix~\ref{App::D0}). This simple analysis already reveals the fundamental difference between the coarse-graining windows: while the $3 \times n$ windows capture the same occupation number for each spin flavor, the $2\times2$ one does not. Consequently, while the first coarse-grained image is uniform for the $3\times n$ window (thus rendering $C_1=0$), the first coarse-grained image using a $2\times 2$ window is not [see, for example, $S_1$ in Fig.~\ref{fig:example_cal}(d)]. This results in non-zero dissimilarities beyond $D_0$ and therefore a finite $C_1$ for the $2\times 2$ window. 

Comparing classical and quantum configurations highlights the central role of quantum fluctuations in the model: whereas classical states fix $C_1=0$ and thus $D_0$ and $C_0$ are constant across all three AFM phases, quantum mean-field snapshots break these constraints. This emphasizes that the dissimilarities and complexity measures act as sensitive probes of quantum fluctuations, particularly displaying distinct signatures across phase boundaries where fluctuations are expected to be stronger.

Beyond simply marking these transitions, the $C_1$ results in Fig.~\ref{fig:c1_12by24} reveal that higher-order structural complexities are deeply sensitive to emergent phases and to how each characteristic spatial periodicity is sampled by different coarse-graining windows. This explains, for example, why $C_1$ obtained with the $3\times 2$ and $3\times 4$ coarse-graining windows show a similar behavior: these two windows are commensurate, and thus probe similar correlations. These findings also show that the AFM patterns have a unit cell larger than $2\times2$, but no larger than $3\times 4$. Additionally, $C_1$ provides a clear diagnostic fingerprint for distinguishing individual magnetic orders: we observe that in the tooth phase $C_1$ raises for the $2\times2$, $3\times2$, and $3\times4$ windows, but vanishes for the $3\times3$ one. At the boundary at $U/t=4.75$, $C_1$ drops for the $2\times2$, $3\times2$, and $3\times4$ windows (by nearly half for the latter two), but rises for the $3\times 3$ window. Inside the zig-zag phase, $C_1$ grows for all windows except $3\times3$, which shows a steady decrease. Finally, the stripe phase is fully captured in $D_0$, and easily identified when using $3\times n$ windows, since $C_1$ vanishes completely.

\begin{table}[tbp!]
  \caption{The sign of $dC_1/dU$ in Fig.~\ref{fig:c1_12by24} across the phase boundaries for different coarse-graining windows.}
  \begin{ruledtabular}
  \begin{tabular}{llccccc}
        Transition & $U/t$  & $2 \times 2$& $3\times 3$ & $3\times 2$ & $3\times 4$\\
     \hline
        Metal $\to$ Tooth  & $3.50$& $+$& $-$ & $+$ & $+$ \\
        Tooth $\to$ Zig-Zag & $ 4.75$& $-$& $+$& $-$& $-$ \\
        Zig-Zag $\to$ Stripe& $ 5.65$ & $-$& $-$& $-$& $-$\\
       \end{tabular}
  \end{ruledtabular}
  \label{table:C1_deriv}
\end{table}

\section{Conclusions}\label{sec::Conc}
In this work, we computed the multiscale structural complexity of spin-resolved mean-field snapshots of an SU(3) FHM at $1/3$-filling in the square lattice as a function of $U/t$. We examined how the geometry of the coarse-graining window influences the dissimilarity $D_0$ and structural complexities $C_0$ and $C_1$, evaluating their ability to identify phase boundaries and ordered phases by benchmarking our results against those from the known mean-field diagram. Our findings validate the structural complexity as a reliable and efficient tool for analyzing SU(N) quantum gas microscope data, providing a theory-independent property that is directly accessible in experiments.

 In this study, we demonstrate that the structural complexity $C_0$ is independent of the coarse-graining window. More significantly, we found that $C_0$ is a robust quantity that provides an experimentally accessible probe to the bipartite entanglement entropy and accurately captures the phase boundaries of the diagram. In addition to $C_0$, $D_0$ and $C_1$ also serve as clear indicators of the metal-to-insulator transition and subsequent transitions to tooth, zig-zag, and stripe phases. These three quantities exhibit extrema in the derivatives at the phase boundaries, reflecting the rapid reorganization of spatial patterns across the lattice as the interaction strength increases. Furthermore, we observed that while $C_0$ and $D_0$ exhibit a similar behavior for all windows used, $C_1$ is highly sensitive to the geometry of the coarse-graining window and phase periodicity. Our results show that, although $C_0$ is dominated by the first coarse-graining step $D_0$ for all windows, quantum fluctuations and correlations beyond the initial coarse-graining window are reflected in $C_1$. Consequently, its derivative $dC_1/dU$ is responsive to both the window geometry and spatial patterns. Moreover, our results show the window-dependent response of $C_1$ can be used as a sensitive probe for distinguishing different magnetic orderings. How to further leverage this sensitivity to elucidate underlying magnetic orders in SU(N$>3$) models and other lattice geometries points to an exciting direction for future research.

\begin{acknowledgments}
This material is based upon work supported by the National Science Foundation under Award No. 2408259. We are grateful to Ehsan Khatami and Richard Scalettar for valuable discussions.
\end{acknowledgments}

\appendix

\section{Post-selection procedure}\label{App::postselection}

The results in the main text are presented after post-selecting the spin-resolved mean-field snapshots using an energy cutoff $\Delta E_\mathrm{cut}/t= 4 \times 10^{-3}$, and for $U/t \in [4.8,5.6]$, after phase filtering. In this Appendix, we quantify the effects of these two filters. 

\begin{figure}[tbp!]
    \centering
    \includegraphics[width=1.0\linewidth]{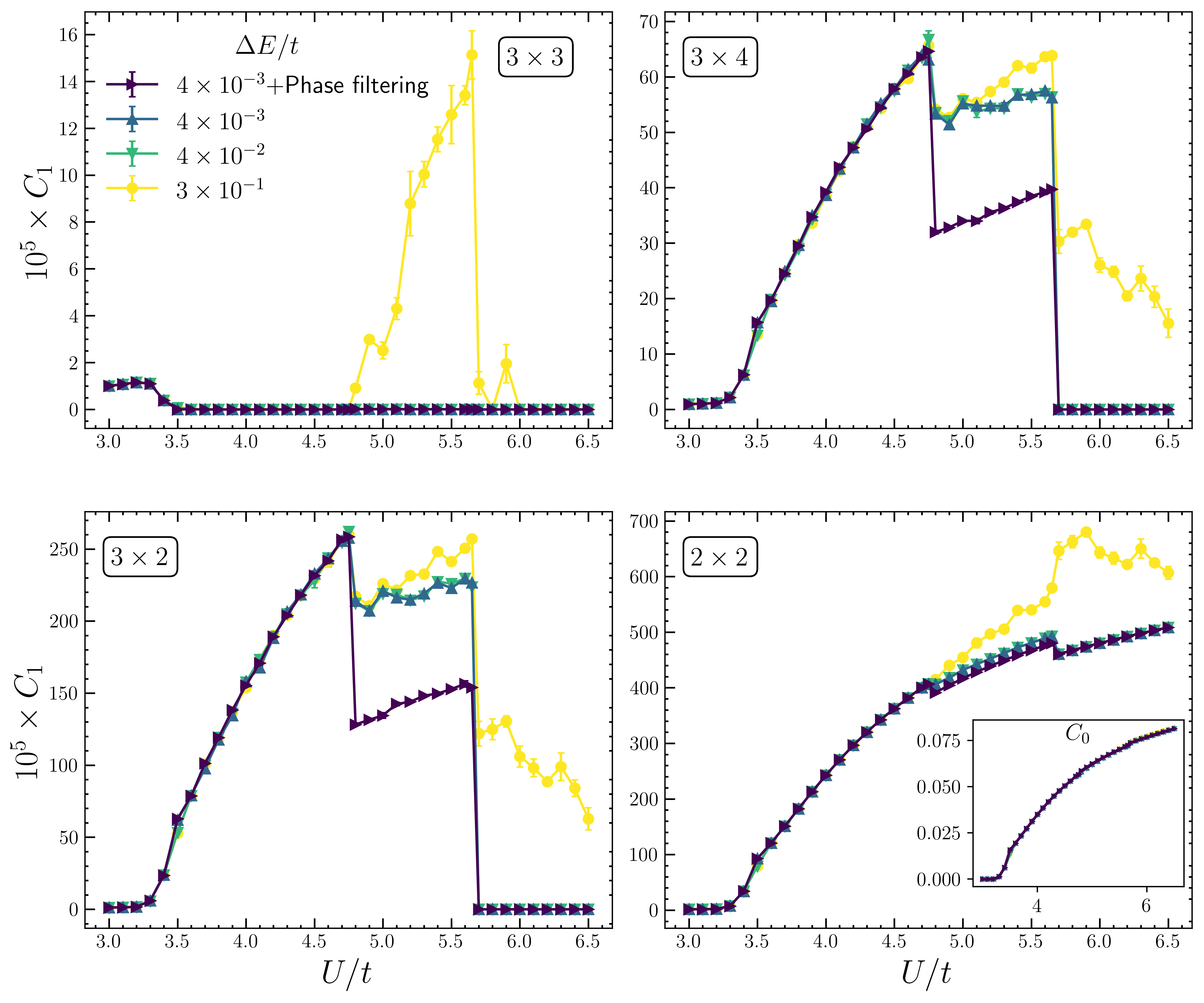}
    \caption{$C_1$ as a function of $U/t$ for different coarse-graining windows and $\Delta E_\mathrm{cut}/t$. Inset: $C_0$ vs $U/t$ for the $2 \times2$ window for the same energy cutoffs.}
    \label{fig:c1_tol}
\end{figure}

In Fig.~\ref{fig:c1_tol} we show $C_1$ and $C_0$ (inset) as functions of $U/t$ for different post-selection criteria. Our results reveal that neither filter has an effect on $C_0$ (only shown for $2\times2$, but conclusions hold for all windows), and that $C_1$ is only sensitive to $\Delta E_\mathrm{cut}/t$ for $U/t>4.75$. The trends with $\Delta E_\mathrm{cut}/t$ are that, as the energy cutoff is made more stringent, (1) $C_1$ gets smaller, and (2) noise is reduced. This is particularly evident for the $3\times 3$ window in the $U/t\in (4.75, 5.65)$ region. For $\Delta E_\mathrm{cut}/t=3\times10^{-1}$, a broad, noisy peak is present, but when a stricter $\Delta E_\mathrm{cut}$ is used, the peak vanishes and $C_1$ flattens close to zero. This additionally suppresses the noise observed at $U/t=5.9$ for the looser cutoff.

This behavior of $C_1$ for the other coarse-graining windows requires further discussion. For the $3\times2$ and $3\times4$ windows, the value of $C_1$ remains almost unaffected in the $U/t\in (4.75, 5.65)$ region as $\Delta E_\mathrm{cut}$ gets stricter. However, within the stripe phase [$U/t\in(5.65,6.5)$], there is a rapid suppression of fluctuations as the energy cutoff gets smaller. 

\begin{figure}[tbp!]
    \centering
    \includegraphics[width=1\linewidth]{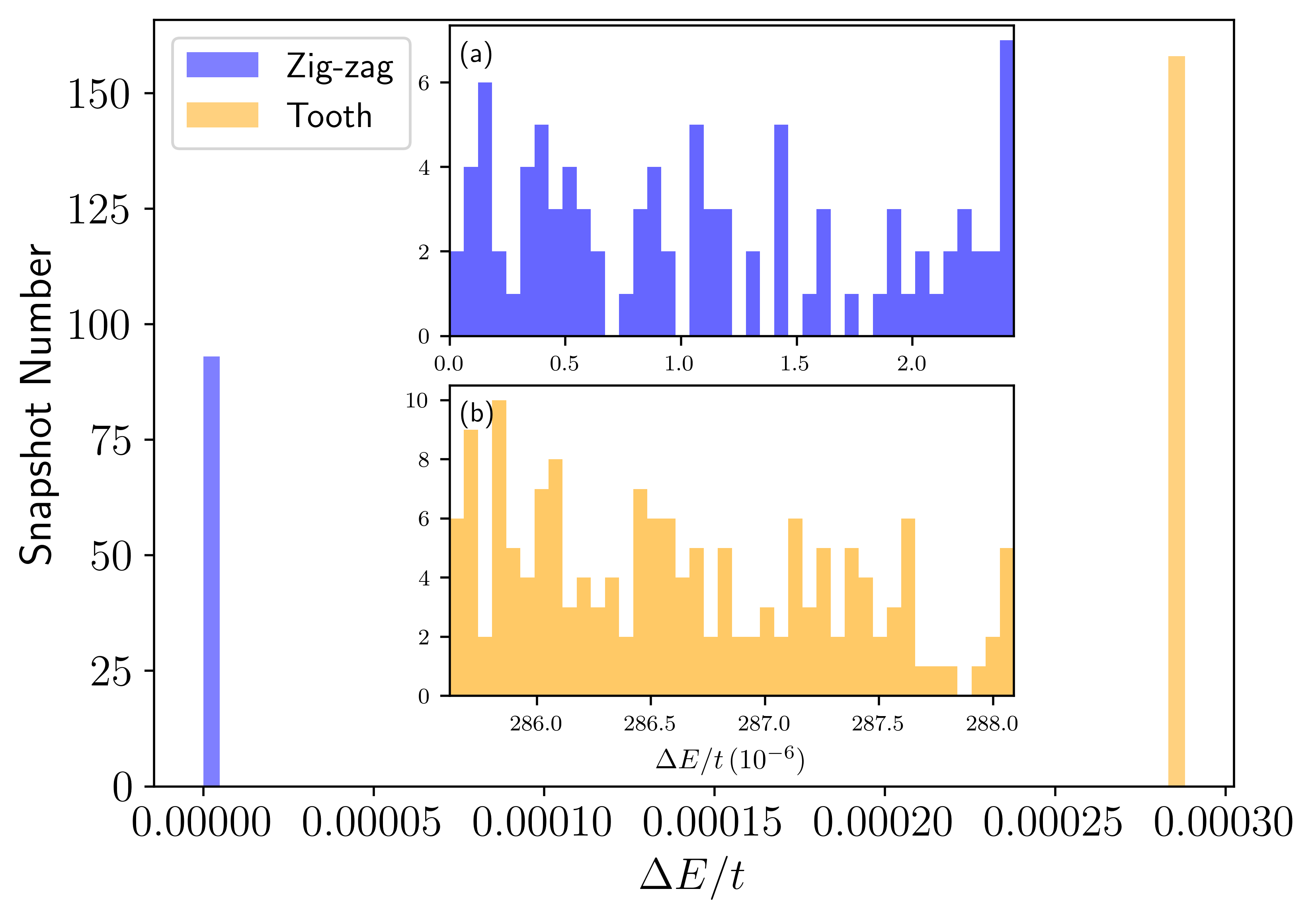}
    \caption{Histogram of the low-energy distribution at $U/t=4.8$. The zig-zag (blue, left bar) and tooth (orange, right bar) phases are separated by a small energy gap $\Delta E\sim 10^{-4}$. Insets: Zoomed-in histograms of the zig-zag (a) and tooth (b) phases. We employed 251 converged energy solutions at this value of $U/t$ before applying point-group symmetries.}
    \label{fig:histogram}
\end{figure}

The weak sensitivity of $C_1$ to $\Delta E_\mathrm{cut}$ in the $U/t\in (4.75, 5.65)$ region is explained by the strong competition between the tooth and zig-zag phases. This is illustrated in Fig.~\ref{fig:histogram}, where these two phases are separated by $\Delta E/t \sim 10^{-4}$. Rather than employing a more stringent cutoff across all values of $U/t$, we phase-filter the data by rejecting snapshots in the higher-energy tooth phase in the competing regime. This procedure reduces noise and decreases the magnitude of $C_1$ without altering its overall trend with $U/t$. Note that an extremely strict energy cutoff across all $U/t$ reduces the number of available snapshots without providing a cleaner signal in $C_1$. In particular, this affects the results in the metallic phase ($U/t<3.50$), where the system is gapless. 

Finally, for the $2\times2$ window, the same trends with $E_\mathrm{cut}/t$ are observed, but the trend with $U/t$ is modified. Beyond reduced noise, $C_1$ decreases rather than increases with larger $U/t$, and exhibits small drops at the phase boundaries. This has important implications in the sign of $dC_1/dU$ across the phase boundaries, and thus demonstrates that the behavior of $C_1$ is strongly influenced by the presence of excited and metastable configurations, highlighting the importance of post-selection.

In summary, our analysis reveals that $C_0$ is insensitive both to variations in $\Delta E_\mathrm{cut}/t$ across three orders of magnitude and to the application of the more stringent phase filter. Furthermore, it also reveals that the jittery signatures in $C_1$ in the unfiltered data are caused by the presence of excited states in the self-consistent mean-field solutions. Therefore, post-selection has two benefits: (1) it allows us to analyze snapshots as close to the ground state as possible without affecting $C_0$. This highlights the robustness of the structural complexity measure $C_0$, validating is use as an efficient and reliable tool for analyzing quantum gas microscopy snapshots to locate phase boundaries. (2) It provides an estimate of the temperatures that need to be reached in experiments with ultracold SU(3) quantum gases in OLs to resolve regions of the phase diagram where strong competition between phases in present. More precisely, when $k_BT \sim \Delta E_\mathrm{cut}$, excited states within that energy gap will contribute to the signal in projective measurements, and this will be accessible to the unfiltered $C_1$ measure.

\section{Density-density correlation function}\label{App:correlation}

\begin{figure}[tbp!]
    \centering
    \includegraphics[width=\linewidth]{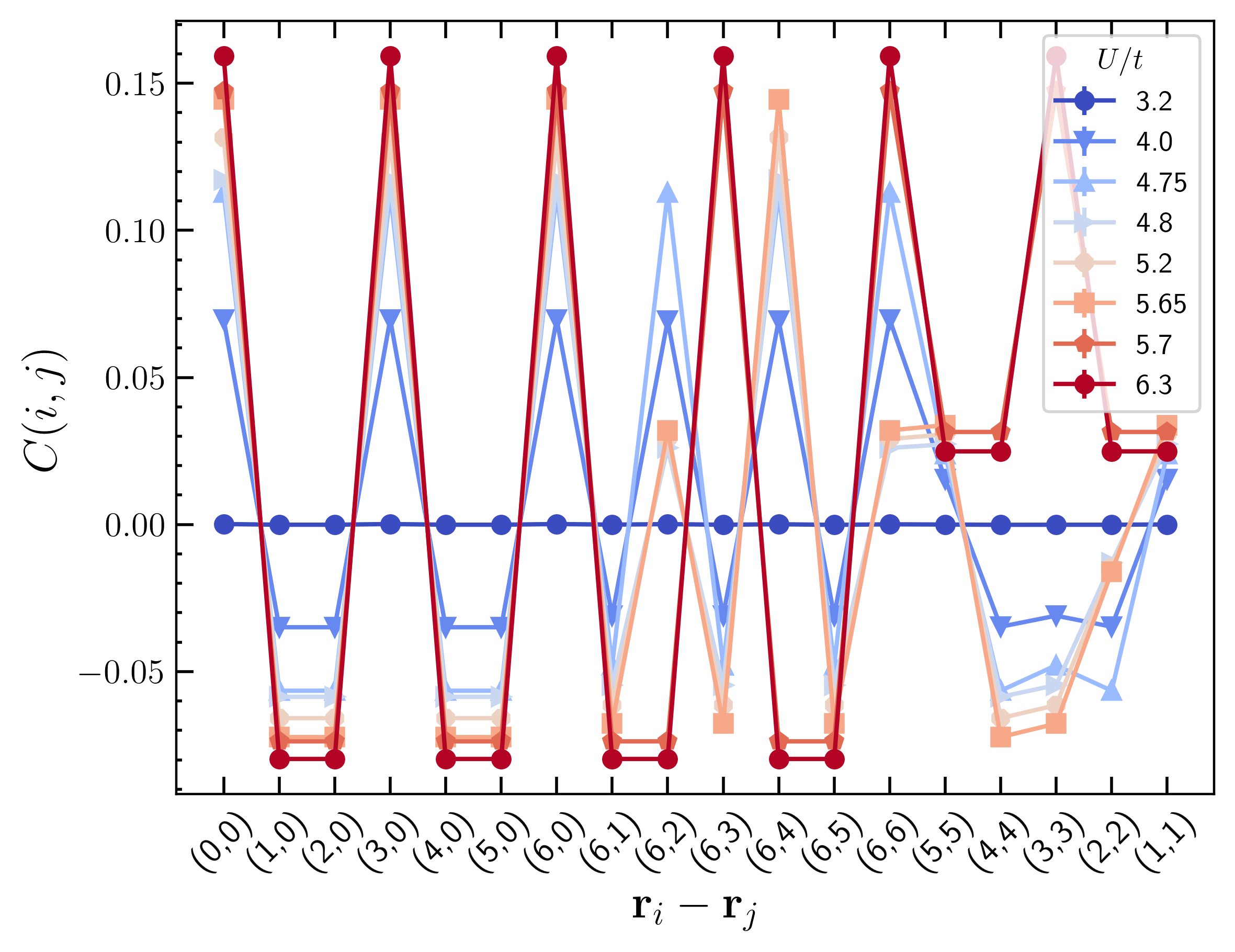}
    \caption{$C(i,j)$ vs $\mathbf{r}_i - \mathbf{r}_j$ for different values of $U/t$.}
    \label{fig:correlation_fun}
\end{figure}

In Fig.~\ref{fig:correlation_fun} we present the correlation function $C(i,j) = \expect{n_{i\sigma}}\expect{n_{j\sigma}} - \expect{n_\sigma}^2$, which measures the strength of the density-density correlations with respect to the uniform density distribution for spin flavor $\sigma$. 

In the metallic phase ($U/t=3.2$), the density distribution is uniform, and $C(i,j)$ vanishes. For $U/t>3.50$, $C(i,j)$ reveals the unit cells of the AFM orderings: For all of them, $C(i,j)$ has a 3-site periodicity in the $\hat{x}$ direction [$(0,0) \to (6,0)$], a 2-site periodicity in the $\hat{y}$ direction [$(6,0) \to (6,6)$] for the tooth and zig-zig phases, and a 3-site periodicity in the $\hat{y}$ and $\hat{x}+\hat{y}$ [$(6,6) \to (0,0)$] directions for the stripe phase. Deep in the AFM phases, i.e. away from the phase boundaries at $U/t=4.75$ and $5.65$, the magnitude of $C(i,j)$ increases monotonically as $U/t$ increases, indicating stronger local moment formation. This explains why, within each phase, $C_0$ and $D_0$ increase as $U/t$ increases. 

The behavior of $C(i,j)$ across the phase boundaries is interesting. When crossing from the tooth ($U/t=4.75$) to the zig-zag ($U/t=4.80$), the most evident change is that although $C(6,2)$ remains positive, its magnitude decreases by more than a factor of 2. On the other hand, when crossing from the zig-zag ($U/t=5.65$) to the zig-zag ($U/t=4.70$), we observe that $C(6,2)$, $C(6,3)$, $C(6,4)$, and $C(3,3)$ reverse sign. 

\FloatBarrier
\FloatBarrier

\section{Expressions for $D_0$ using the $2\times2$ window for classical snapshots}\label{App::D0}

To simplify reading the following equations, in this appendix we follow the notation used in Ref.~\cite{IbarraSC2024}, and use two indices $ij$ to label a site in an image. We consider a two-dimensional image with linear dimension $L$ and $N= L\times L$ sites. On each site, given by the coordinate pair $ij$, the value of the image is $s_{ij}$. The first term $D_0$ for a $2\times2$ window is given by
\begin{widetext}
    \begin{equation}
         D_0 = \frac{1}{4N}  \bigg\vert  \sum_{i,j=1}^{L/2} \bigg[ s_{2i-1,2j-1}(s_{2i-1,2j} + s_{2i,2j-1}) + s_{2i,2j-1}(s_{2i-1,2j} + s_{2i,2j}) + s_{2i,2j}(s_{2i-1,2j-1} + s_{2i-1,2j}) \bigg] - \frac{3}{2} \sum_{i,j=1}^L s_{ij}^2 \bigg\vert
    \end{equation}
\end{widetext}
and it captures all correlations within the coarse-graining window, i.e. on-site, nearest and next-nearest neighbors.

In the case of spin-resolved snapshots the images correspond to $s_{ij} = n_{ij,\sigma}$. So, $D_0$ is given by,
\begin{equation}
    D_0 = \frac{1}{4}  \bigg\vert  \expect{n_{\sigma}n_{\sigma}}_{nn} +\frac{1}{2} \expect{n_{\sigma}n_{\sigma}}_{nnn}
   - \frac{3}{2} \expect{n_{\sigma}} \bigg\vert,
\end{equation}
where we exploited the translational and rotational symmetry. Here $\expect{n_\sigma}$ is the density for spin flavor $\sigma$, and $\expect{n_{\sigma}n_{\sigma}}_{nn}$  and $\expect{n_{\sigma}n_{\sigma}}_{nnn}$ are shorthand for the nearest-neighbor and next-nearest-neighbor density-density correlations. For the classical configurations of the tooth, zig-zag, and stripe phases, $\expect{n_{\sigma}n_{\sigma}}_{nn}=0$ because there are no two adjacent sites occupied by the same spin flavor. Additionally, $\expect{n_{\sigma}n_{\sigma}}_{nnn}= 1/6$, because there are only 3 distinct patterns within the $2\times2$ windows, and in those 5 out of the 6 nearest-neighbor bonds vanish. Thus, at $1/3$-filling
\begin{equation}
    D_0 = \frac{1}{4} \bigg\vert \frac{1}{2}\frac{1}{6} - \frac{3}{2}\frac{1}{3} \bigg\vert = \frac{5}{48}
\end{equation}
for the classical configurations of the three AFM phases. 

An analytical derivation of $D_{k>0}$ becomes untractable. However, we numerically observe that $C_0 \to 1/9$ as $k\to \infty$, so we conclude that $C_1 = 1/144$.

\bibliography{SU3}

\end{document}